# Crystallization of heavy fermions via epitaxial strain in spinel LiV$_2$O$_4$ thin film

U. Niemann[1,2*], Y.-M. Wu[1*], R. Oka[1,2], D. Hirai[2], Y. Wang[1], Y. E. Suyolcu[1], M. Kim[1], P. A. van Aken[1], and H. Takagi[1,2,3]

[1]Max Plank Institute for Solid State Research, Heisenbergstraße 1, 70569 Stuttgart, Germany

[2]Department of Physics, University of Tokyo, Hongo 7-3-1, Bunkyo-ku, Tokyo 113-0033, Japan

[3]Insitute for Functional Matter and Quantum Technologies, University of Stuttgart, Pfaffenwaldring 57, 70550 Stuttgart, Germany.

[*]These authors contributed equally to this work.

The mixed-valent spinel $LiV_2O_4$ is known as the first oxide heavy-fermion system. There is a general consensus that a subtle interplay of charge, spin, and orbital degrees of freedom of correlated electrons plays a crucial role in the enhancement of quasi-particle mass, but the specific mechanism has remained yet elusive. A charge-ordering (CO) instability of $V^{3+}$ and $V^{4+}$ ions that is geometrically frustrated by the V pyrochlore sublattice from forming a long-range CO down to $T = 0$ K has been proposed as a prime candidate for the mechanism. To uncover the hidden CO instability, we applied epitaxial strain from a substrate on single-crystalline thin films of $LiV_2O_4$. Here we show a strain-induced crystallization of heavy fermions in a $LiV_2O_4$ film on MgO, where a charge-ordered insulator comprising of a stack of $V^{3+}$ and $V^{4+}$ layers along [001], the historical Verwey-type ordering, is stabilized by the in-plane tensile and out-of-plane compressive strains from the substrate. Our discovery of the [001] Verwey-type CO, together with previous realizations of a distinct [111] CO, evidence the close proximity of the heavy-fermion state to degenerate CO states mirroring the geometrical frustration of the pyrochlore lattice, which supports the CO instability scenario for the mechanism behind the heavy-fermion formation.

The mixed-valent vanadate $LiV_2O_4$ is one of the few metallic spinel oxides and shows a heavy-fermion ground state[1,2]. The origin of mass enhancement of quasi-particles has been a subject of long debates, which has attracted considerable interest both theoretically and experimentally over more than two decades[3-7]. The proposed mechanisms so far include a Kondo mechanism as in $4f$ inter-metallics[3,4], formation of itinerant quantum spin liquid[7], and close proximity to a correlated insulator, such as an orbital-dependent Mott state[5], and a charge-ordered state[6].

$LiV_2O_4$ crystallizes in a cubic spinel structure[8], as shown in Fig. 1a. Octahedrally coordinated V ions are all equivalent and form a pyrochlore sublattice, which consists of a corner-shared network of V tetrahedra and is known to give rise to strong geometrical frustration. The formal valence of V ions is 3.5+ with a 1:1 ratio of $V^{3+}$ and $V^{4+}$. In the electronic band structure, 1.5 $d$-electrons per V ion are accommodated in the $t_{2g}$ band-manifolds. Due to the trigonal distortion of $VO_6$ octahedra, characterized by a spinel $u$-parameter of ~0.26, the $t_{2g}$ band-manifolds split into a narrow band with $a_{1g}$ orbital character and broad bands with doubly-degenerate $e_g$ orbital character. In the presence of a sizable on-site Coulomb interaction, $U$, the narrow $a_{1g}$ band is stabilized with respect to the broad $e_g$ bands and discussed to accommodate ~1 localized electron[3]. The remaining ~0.5 electrons occupy the broad $e_g$ bands with itinerant character.

Kondo coupling between the itinerant $e_g$ electrons and the $a_{1g}$ local moments was suggested as the origin of the heavy mass as in canonical $4f$ heavy-fermion systems[3]. However, it is not obvious to justify a sizable Kondo coupling in $LiV_2O_4$ (ref. [9]). Alternative models were proposed, emphasizing the importance of geometric frustration inherent to the pyrochlore lattice, which suppresses a long-range ordering of charge, spin, and orbitals. The geometrical frustration of the pyrochlore lattice has been long known to give rise to a macroscopic degeneracy of states, since Pauling proposed his idea of the ice rule and the residual ice-rule entropy[10]. The idea was later extended to the cases for spins and 1:1 mixed charges by Anderson[11], who argued that the presence of a macroscopic degeneracy of spin- and charge-ordered states, satisfying the Anderson condition equivalent to the ice rule, prevents the selection of a specific spin/charge-ordered state. If the system is critically close to a charge-ordered insulator or an orbital-dependent Mott insulator, a strongly renormalized, highly entropic metallic state might be anticipated if no charge/orbital ordering is present[5,6].

Previous studies have shown that $LiV_2O_4$ undergoes a metal-to-insulator transition under hydrostatic pressure, resulting in CO along the [111] direction accompanied by orbital ordering and spin-singlet molecule formation[12-16]. The presence of the [111] charge-ordered state may indicate the presence of other "hidden" COs close in energy, manifesting the macroscopic degeneracy of charge-ordered states originating from the geometrical frustration. Out of a pool of degenerate COs, a unique ground state with CO distinct from the [111] ordering may emerge via an external perturbation, for example, lattice deformation that strongly couples with a specific CO.

We have explored such hidden COs by applying epitaxial strain to thin-film structures of spinel $LiV_2O_4$. Epitaxial strain from substrate has been successfully employed to control the spin/charge/orbital state of correlated oxides[17-22]. While Li-containing spinel oxides have been shown to epitaxially grow on various substrates[23,24], two substrates, $SrTiO_3$ (STO) and MgO, were chosen for fabricating $LiV_2O_4$ thin films in this study. As shown in Table 1, the cubic lattice constant, $a_{LVO}$, of bulk $LiV_2O_4$ is 8.24 Å[25]. Those of the substrate materials, STO and MgO, are 3.91 and 4.21 Å, amounting to lattice mismatches relative to $a_{LVO}/2$ of −5.1 % (compressive) and +2.2 % (tensile), respectively. Given the large lattice mismatch, a $LiV_2O_4$ film on STO substrate is fully relaxed and shows heavy-fermion behavior as in bulk $LiV_2O_4$, whereas given the smaller lattice mismatch, a $LiV_2O_4$ film on MgO is epitaxially strained and found to be a charge-ordered insulator comprising of $V^{3+}$ and $V^{4+}$ layers stacked in parallel to the substrate plane.

Single-crystalline thin films on (001) STO and MgO substrates were grown by pulsed laser deposition (PLD). See Methods for details of film growth. Scanning transmission electron microscopy high-angle annular dark-field (STEM-HAADF) images with incident beam parallel to the [110] direction are shown in Figs. 1c (STO) and 1d (MgO), which confirm atomically uniform LiV$_2$O$_4$ films grown along the [001] direction on both substrates. An alternating stack of atomic layers consisting of V chains running along the [110] direction (parallel to the film surface and to the beam) and [1$\bar{1}$0] direction (parallel to the film surface and perpendicular to the beam) can be seen clearly, which corresponds to the schematic picture of the atomic arrangement in Fig. 1b.

We observe contrasted growths between the films on the two substrates, namely, a relaxed growth on STO with a large mismatch and an epitaxially strained growth on MgO with a moderate mismatch. The X-ray $\theta$-$2\theta$ scans for the LiV$_2$O$_4$ thin films (see Figs. S1 and S2 in Supplementary Information) indicate that while the out-of-plane lattice constant, $c$, of the film on STO (8.23 Å) is essentially the same as the bulk lattice constant $a_{LVO}$, that of the film on MgO (8.13 Å) is much smaller than $a_{LVO}$. These observations strongly suggest a fully relaxed film growth on STO and a pseudomorphic growth on MgO. The X-ray reciprocal space mappings (RSM), shown in Figs. 1e and 1f, indeed indicate that the in-plane lattice constant of the film on STO is identical to $a_{LVO}$ and not locked to the substrate. In contrast, that of the film on MgO is clearly locked to the substrate, meaning the LiV$_2$O$_4$ film is subject to an in-plane tensile strain of 2.2 %. The out-of-plane lattice constant is contracted with respect to the bulk by 1.3 %, which can be naturally understood as the Poisson effect[26]. Note that for the LiV$_2$O$_4$ thin film on STO, we observe periodic misfit dislocations in STEM-HAADF images (Figs. 1c, 1d and S3) consistent with the fully relaxed film growth on STO, whereas misfit dislocations are absent for the LiV$_2$O$_4$ thin film on MgO.

The electronic transport properties of the two films on STO and MgO are distinct, which indicates that the ground state of LiV$_2$O$_4$ is successfully controlled via strain. As seen in Fig. 2a, the temperature-dependent resistivity, $\rho(T)$, for the relaxed film on STO is metallic and well reproduces those of bulk single crystals reported so far[2]. The magnitude of the resistivity at room temperature is 800-900 $\mu\Omega$cm, in excellent agreement with those reported for bulk single crystals[2]. A knee-like structure is observed at around $T^* = 25$ K, below which $\rho(T)$ shows a more pronounced decrease by lowering the temperature. $T^*$

has been interpreted as a coherence temperature for the formation of quasiparticles. In the low-temperature limit below $T \sim 1$ K, a $T^2$-dependence of $\rho(T)$ is observed, consistent with a Fermi liquid ground state. The coefficient of the $T^2$ term, $A \sim 2.4$ $\mu\Omega$cmK$^{-2}$, is close to the reported value of $A \sim 2.0$ $\mu\Omega$cmK$^{-2}$ in bulk samples[2]. The $A$ value satisfies the Kadowaki-Woods relationship[27] and is consistent with the reported large electronic specific heat coefficient, $\gamma \sim 400$ mJ/molK$^2$ (refs. [1,2,28]), which has been taken as evidence for the heavy-fermion ground state[2].

At high temperatures well above $T^*$, $\rho(T)$ shows almost $T$-linear behavior. This seemingly metallic behavior persists above the Ioffe-Regel limit, $\rho_{IR} \sim 600$ $\mu\Omega$cm, implying incoherent transport[2]. In the corresponding high-temperature region, the Drude peak in the optical conductivity, which represents a coherent transport and is well-defined at low temperatures below $T^*$, was indeed found to fade out, indicative of a collapse of coherent transport[6]. The spectral weight of the Drude peak well below $T^*$ is transferred to high energies up to a few eV (ref. [6]) in the high-temperature incoherent-transport regime. The large energy scale of the spectral-weight transfer demonstrates the vital role of electron correlations in the formation of heavy quasi-particles and was argued to represent the CO instability[6].

The epitaxially strained film on MgO, in stark contrast to the relaxed film on STO, shows insulating behavior in $\rho(T)$ below room temperature, as seen in Fig. 2b. The maximum activation energy estimated from the Arrhenius plot (inset of Fig. 2b) is ~50 meV, indicating that the ground state is an insulator with a charge gap of the order of 100 meV. The expected charge gap is comparable to the charge gap estimated from the optical conductivity of the charge ordered/dimerized Magneli phase V$_4$O$_7$ (ref. [29]). As mentioned above, the quality of the lattice on the atomic level seen in the STEM image (Figs. 1c and 1d) for the film on MgO is at least comparable to that on STO. In addition, the film on MgO shows a much smaller full width at half maximum (FWHM) in the XRD rocking curve and hence disorder as compared to that on STO (Figs. S1 and S2), very likely reflecting its better lattice match with the substrate. The similar or even better quality of the film on MgO as compared with that on STO indicates that its insulating behavior does not originate from Anderson localization due to increased disorders but an intrinsic electronic transition induced by the strain from the substrate. The natural candidate behind this heavy-fermion-to-insulator transition is CO of $V^{3+}$ and $V^{4+}$.

The signature of CO in the insulating strained film on MgO was explored by conducting a spatially resolved electron energy-loss spectroscopy (EELS) combined with STEM. It is known that the intensity ratio of metal $L$-edge peaks ($L_2$ and $L_3$), $I(L_2)/I(L_3)$, which we write as the $L_2/L_3$ ratio for simplicity, varies in transition-metal oxides as a function of $d$-electron filling[30]. In particular, the $L_2/L_3$ ratio is appreciably different between $V^{3+}$ and $V^{4+}$, ~0.83 for $V^{3+}$($3d^2$) to ~1.02 for $V^{4+}$($3d^1$) (refs. [31,32]), which serves as a good measure of the valence state of V ions (Fig. S4). Spatially resolved EELS spectra were measured on (110) STEM images for both the relaxed metallic film on STO (Fig. 3a) and the strained insulating film on MgO (Fig. 3c). A coarse line scan was first performed with a step size of 1.8 Å along the direction perpendicular to the film surface indicated by the white arrow in Figs. 3a and 3c. For the coarse scan, we employ an averaging over a width of 28.2 Å perpendicular to the scan direction (indicated by the vertical bar to the arrow) to minimize beam damage to the film (Fig. S5). We note that the direction of averaging is parallel to the film surface and the [$1\bar{1}0$] chain direction. The spatial variation of the $L_2/L_3$ ratio obtained from each EELS spectrum is shown in Figs. 3b and 3d. The averaged $L_2/L_3$ ratios over the measured distance, shown by the horizontal broken line in Figs. 3b and 3d, are almost identical between the strained film on MgO and the relaxed film on STO, ~0.9, which is in between 0.83 for $V^{3+}$ and 1.02 for $V^{4+}$ (Fig. S4) and fully consistent with the average valence of V ions ~3.5+. The variation of the $L_2/L_3$ ratio from the measured point to point, however, is quite distinct between the two films. While the distribution of the $L_2/L_3$ ratio for the relaxed metallic film on STO is concentrated around 0.9, that for the strained insulating film on MgO scatters widely from 0.8 to 1.05 and shows two peaks at ~0.84 and ~0.98. The former is consistent with a spatially homogeneous distribution of a $V^{3.5+}$ expected for the metallic state, as in bulk crystals. The latter strongly suggests a spatially inhomogeneous distribution of $V^{3+}$ and $V^{4+}$, indicative of a CO. The fact that the $L_2/L_3$ ratio takes values close to those expected for $V^{3+}$ or $V^{4+}$ even after averaging over 28.2 Å perpendicular to [001] scan direction indicates the homogeneous distribution of either $V^{3+}$ or $V^{4+}$ along the [1$\bar{1}$0] chain direction for a given spot of the measurement and, therefore, an ordering of $V^{3+}$ and $V^{4+}$ layers along [001] direction.

To verify the suggested [001] ordering of $V^{3+}$ and $V^{4+}$ layers, atom-specific EELS measurements were conducted on the V atoms lining up along the two directions, [1$\bar{1}$2] and [$\bar{1}$12], as shown in the insets in Figs. 4a and 4d, respectively. Only four atoms were measured in each scan, due to the electron damage arising from a much higher intensity of beam than that in the coarse scan. The high intensity was necessary to ensure the required atomic resolution with a reasonable signal-to-noise ratio. The $L_2/L_3$ ratio shows oscillatory behavior both along the [1$\bar{1}$2] and [$\bar{1}$12] scans. We find that the V ions in the layer of the [1$\bar{1}$0] chains (colored blue in Fig. 4) show $L_2/L_3 \sim 1$ corresponding to a 4+ valence and that the neighboring V ions in the [110] chains (colored yellow in Fig. 4) show $L_2/L_3 \sim 0.85$ corresponding to a 3+ valence. These results confirm stacking of a $V^{3+}$ layer with a [110] chain and a $V^{4+}$ layer with a [1$\bar{1}$0] chain alternating along [001], perpendicular to the film plane, as schematically depicted in Fig. 5a. The indicated CO pattern of $V^{3+}$ and $V^{4+}$ is nothing but that originally proposed by Verwey for magnetite $Fe_3O_4$ (ref. [33]). Heavy fermions in $LiV_2O_4$ crystallize via epitaxial strain.

The stack of [110] $V^{3+}$ and [1$\bar{1}$0] $V^{4+}$ chains, orthogonal to each other, makes the two chains inequivalent and results in at least an orthorhombic symmetry. Two kinds of domains would be formed on the four-fold symmetric MgO (001) surface, domain I with [110] $V^{3+}$ and [1$\bar{1}$0] $V^{4+}$ chains and domain II with [1$\bar{1}$0] $V^{3+}$ and [110] $V^{4+}$ chains. Within the measurements repeated a few times, we always observed $V^{3+}$ in the [110] chains and $V^{4+}$ in the [1$\bar{1}$0] chains, namely, domain I. Domain I might be stabilized in the sample thinning process for the STEM measurement.

The stabilization of a [001] Verwey-type CO, a distinct class of CO from the [111] ordering observed under hydrostatic pressure[12-16], indicates the presence of at least two distinct, energetically almost degenerate charge-ordered states in the close vicinity to the heavy-fermion state, which we argue to reflect the geometrical frustration in distributing $V^{3+}$ and $V^{4+}$ ions in a 1:1 ratio. The suppression of the occurrence of a long-range CO due to the degeneracy may allow the persistence of the metallic state with extensive short-range CO fluctuations and the diverging enhancement of quasi-particle mass, if the system is critically close to the border of a metal-to-charge-ordered-insulator transition. Our results are fully consistent with such a scenario based on the CO instability and the geometrical frustration.

We argue that the [001] Verwey-type CO is selected by the anisotropic epitaxial strain from the substrate, instead of the [111] ordering observed under hydrostatic pressure[12-16]. Considering charge neutrality, each $V^{3+}$ and $V^{4+}$ layer has a spatial charge of $-0.5e$ and $+0.5e$ per V ions respectively, forming a stack of negatively and positively charged layers parallel to the substrate plane, as schematically shown in Fig. 5b. The intra- and inter-layer Coulomb interactions are repulsive and attractive, respectively, which should favor an in-plane lattice expansion and an out-of-plane lattice contraction. This perfectly matches with the in-plane tensile and out-of-plane compressive epitaxial strains from the interface with the MgO substrate.

The [111] CO of $LiV_2O_4$ under hydrostatic pressure[14], as well as the [111] ordering in the related compound $AlV_2O_4$ (ref. [34]), are accompanied by orbital ordering and molecular-orbital formation. In magnetite $Fe_3O_4$, a molecular orbital called a trimeron formed through an orbital ordering has been identified in the charge-ordered state[14,35]. These orbital orderings and molecular-orbital formations have been believed to play a crucial role in selecting the respective charge-ordered states, giving rise to a complex ordering pattern of charge. The [001] Verwey-type CO in the strained $LiV_2O_4$ film unveiled in this study appears to have a much simpler ordering pattern than the other relevant cases. We do not see apparent dimerization/molecular orbital of V atoms in the STEM images. The in-plane tensile and out-of-plane compressive strains from the substrate are uniform in this specific case. The trigger by such uniform strain, rather than by a local lattice distortion such as dimerization/molecular orbital formation, might promote the selection of such a simple ordering pattern. We note however that even without apparent dimerization/molecular orbital formation, orbital degrees of freedom could be still relevant in realizing the specific CO state observed in this study. If, for example, the orbital states for $V^{4+}$ and $V^{3+}$ are simply $a_{1g}^1 e_g^0$ and $a_{1g}^1 e_g^1$, the displacement of surrounding O anions should take place to destabilize and stabilize the $e_g$ state for $V^{4+}$ and $V^{3+}$, respectively. The exploration of such orbital physics is worthy of future exploration, which requires detailed information on the local crystal structure, including oxygen ions.

Thin-film technology has been successfully utilized in controlling the competing spin/charge/orbital orders of correlated transition-metal oxides. Fully taking the merit of the success, we have discovered a crystallization of heavy fermions into the [001] Verwey-type CO in the strained $LiV_2O_4$ film on the MgO substrate, where the tensile strain in the (001) plane and the compressive strain along [001] stabilize the Verwey CO instead of the previously observed [111] CO. The discovery, together with the observation of the [111] ordering under hydrostatic pressure, establishes the presence of the CO instability in heavy-fermion $LiV_2O_4$, and more importantly, the degeneracy of the competing charge-ordered states, a manifestation of strong geometrical frustration inherent to the pyrochlore sublattice of spinel. This may favor the CO instability scenario as the mechanism for the formation of heavy quasi-particles in the long-studied spinel oxide $LiV_2O_4$. It hints at various exciting directions of research as an extension. Using the degeneracy of different CO patterns, for example, we may be able to discover/tune other distinct and hidden CO patterns by applying strain along many different crystallographic directions using a film structure and/or uniaxial pressure.

## Methods

### Thin film growth

Single-crystalline thin films of $LiV_2O_4$ were grown on (001) $SrTiO_3$ (STO) substrate and on MgO substrate by a pulsed laser deposition (PLD) technique. A mixed-phase Li-V-O pellet was used as a target material. $LiCO_3$ and $V_2O_5$ were mixed with a molar ratio of 1.75:1, followed by pelletizing and sintering at 500 °C for 72 hours. The pulverization, pelletization and sintering processes under the same condition were repeated a few times. The target was ablated with a pulsed KrF excimer laser ($\lambda = 248$ nm) at 5 Hz with a flux density of 3.3 $Jcm^{-2}$. The substrate temperature was kept at 450 °C during the growth. The typical film thickness was 20-40 nm, which corresponds to 24-48 unit cells.

### X-ray and transport characterization

A high-resolution X-ray diffractometer (*Smartlab, Rigaku*) with a Cu $K\alpha_1$ source was used for the structural characterization of the grown films. The film thickness was calibrated from Kiessig oscillations in X-ray reflectivity (see Figs. S1 and S2 in Supplementary Information). The temperature-dependent resistivity was measured using a Physical Property Measurement System (*Quantum Design*) in a conventional four-probe configuration.

### STEM and EELS measurements

The TEM sample preparation includes a mechanical grinding (down to ∼10 μm), a tripod wedge polishing (with an angle of ∼1.5°), and a double-sided argon-ion milling. Isopropyl alcohol was used to avoid any contact of the specimen with water during the grinding and the polishing. For the argon-ion thinning, a Gatan precision ion polishing system II (PIPS II, Model 695) was used at liquid nitrogen temperature. Right before the experiment, the samples were treated in a Fischione plasma cleaner in a 75% argon - 25% oxygen mixture.

For the STEM-EELS experiments, a probe aberration-corrected JEOL JEM-ARM200F STEM equipped with a cold field-emission electron source, a probe Cs-corrector (DCOR, CEOS GmbH), a Gatan GIF Quantum ERS spectrometer, and a Gatan K2 direct electron detector were used at 200 kV. The STEM imaging and the EELS analyses were performed at probe-convergence-semiangles of 20 and 28 mrad, resulting in probe sizes of 0.8 and 1.0 Å, respectively. The collection angle for HAADF imaging ranged from 110 to 270 mrad. A collection semiangle of 111 mrad was used for the EELS investigations. A 0.1 eV/ch dispersion with an effective energy resolution of ~0.7 eV was chosen particularly for the V $L_{2,3}$ white lines to quantify the V $L_2/L_3$ intensity ratio with a sufficiently high signal-to-noise ratio. ~1.8 Å scan step size, 28.2 Å averaging width and 10 ms per pixel dwell time was used to probe the overall valence distribution of two films in Fig. 3. ~0.5 Å scan step size and 5 ms per pixel dwell time was used to resolve the atomic-resolution V valence in Fig. 4. Further details of the data processing are given in Supplementary Information.

**Data availability**

The data that support the findings of this work are available from the corresponding authors upon request.

**Acknowledgements**

We thank J. Bruin, D. Huang, A. Rost and A. Yaresko for discussions, and G. Cristiani, M. Dueller, G. Logvenov, S. Prill-Diemer and B. Stuhlhofer for experimental support. This work was partly supported by a Grand-in-Aid for Promotion of Scientic Research (KAKENHI), Ministry of Education, Sports and Culture of Japan (Grant No. JP22H01180). This project has received funding from the European Union's Horizon 2020 research and innovation programme under grant agreement No. 823717 – ESTEEM3.

**Author contributions**

U.N., D.H., R.O. and M.K grew the films and carried out the XRD and transport measurements. Y.-M.W., Y.W., Y.E.S, P.vA conducted the STEM and EELS measurements and analyzed the data. H.T., U.N. and Y.-M.W discussed the implication of the data. U.N., Y.-M.W, M.K. and H.T. wrote the manuscript with inputs from all the other authors. H.T. conceived the project.

**Competing interests**

The authors declare no competing interests.

**Additional information**

Supplementary information

Supplementary Note 1, Figures S1-S5.

**Correspondence**

Correspondence to Minu Kim (minukim@fkf.mpg.de) and Hidenori Takagi (h.takagi@fkf.mpg.de).

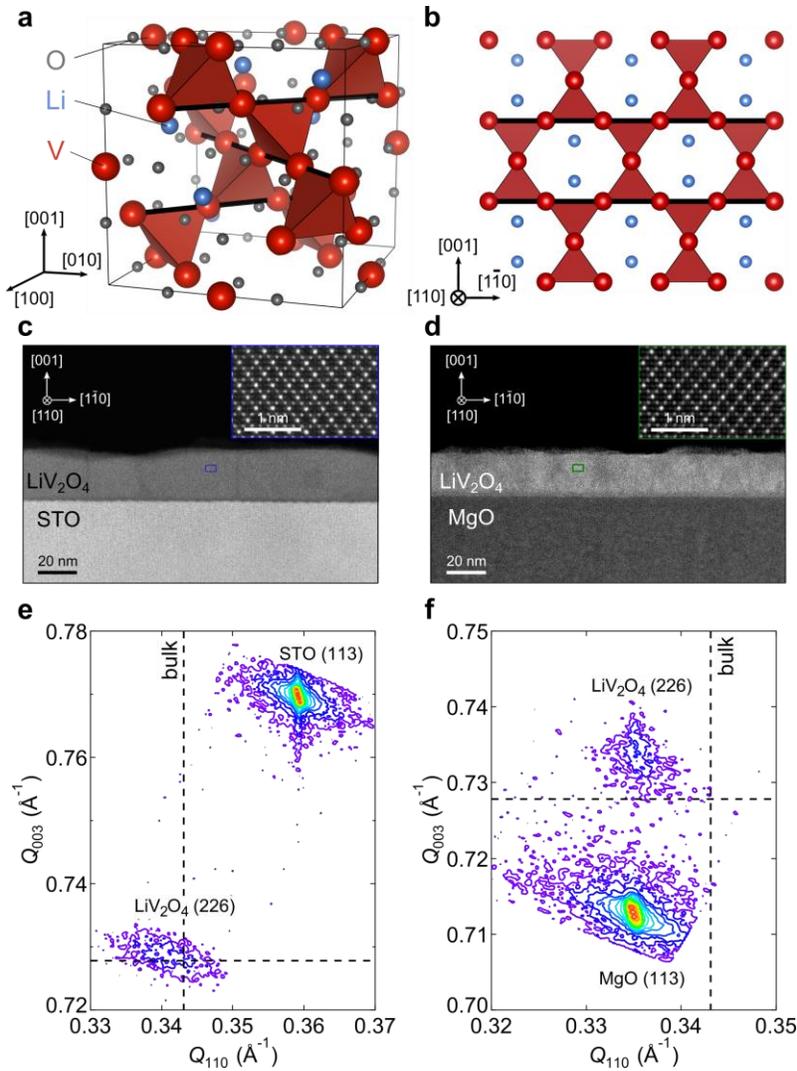

**Fig. 1 | Structural characterization of LiV$_2$O$_4$ films grown on STO and MgO. a,** The crystal structure of the spinel LiV$_2$O$_4$. Red, gray and blue balls indicate V, O and Li atoms, respectively. **b,** The (110) view of the crystal structure in **a**. For simplicity, O ions are not included. The black lines indicate the V chains orthogonal to each other along [1$\bar{1}$0] and [110] directions. **c,d,** STEM-HAADF images of the LiV$_2$O$_4$ films on STO and MgO substrates, respectively. The insets are zoomed-in views of the blue and green rectangles in the images. **e,f,** X-ray reciprocal space mapping around the (226) Bragg peak of the films on STO and MgO, respectively. $Q_{003}$ and $Q_{110}$ are the inverse of lattice spacing along [003] and [110], respectively. The intersection of the black dashed lines denotes the position of the (226) peak expected for bulk LiV$_2$O$_4$ (ref. [25]).

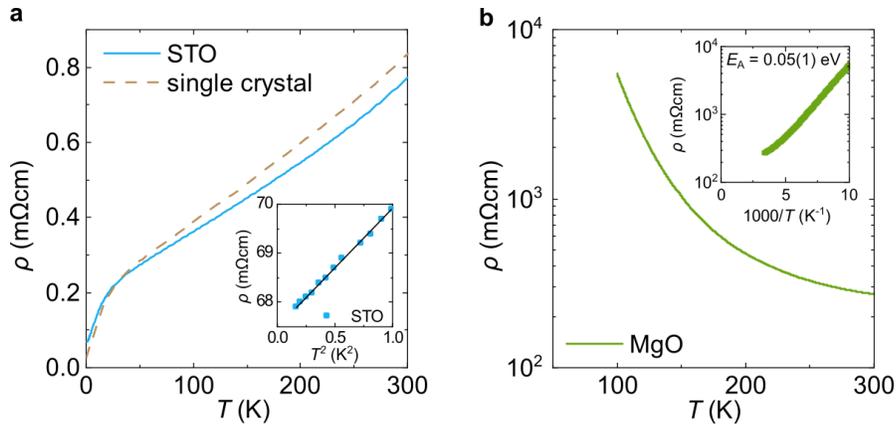

**Fig. 2 | Temperature-dependent resistivity $\rho(T)$ of unstrained and strained LiV$_2$O$_4$ films**. **a,** $\rho(T)$ for the relaxed film on STO (blue) shows metallic behavior, which well reproduces that of bulk single crystal (orange)[2]. The $\rho$-$T^2$ plot below 1 K in the inset shows the $T^2$-dependence of $\rho(T)$ at low temperatures, indicating a Fermi-liquid ground state. **b,** $\rho(T)$ for the strained film on MgO (green) shows insulating behavior with an activation energy $E_A$ of ~50 meV, estimated from the Arrhenius plot of $\rho(T)$ in the inset.

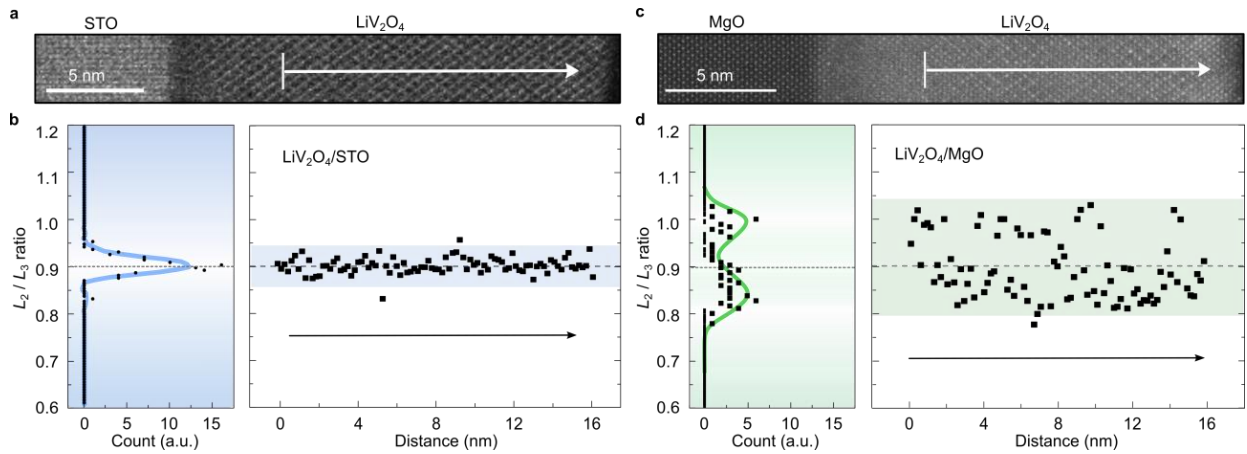

**Fig. 3 | The spatial distribution of V valence states in LiV₂O₄ films on STO and MgO substrates. a,c,** STEM-HAADF images of the unstrained and strained LiV₂O₄ films on STO and MgO, respectively. The spatial variation of EELS spectrum was measured by scanning the white line perpendicular to the film surface with a step size of a 1.8 Å. The vertical line to the white arrow indicates an averaging width of 28.2 Å for each EELS measurement. **b,d,** The spatial profiles of the $L_2/L_3$ intensity ratio for the films on STO (**b**) and MgO (**d**), respectively. The black arrows indicate the direction of the line-scan. The $L_2/L_3$ ratio is defined as the ratio of maximum intensities of the V-$L_3$ and V-$L_2$ white lines. The left panel of b and d shows the statistical distribution of the $L_2/L_3$ ratio. A homogeneous charge distribution can be seen for the unstrained LiV₂O₄ shown in **b**. In contrast, in the strained LiV₂O₄ shown in **d**, two peaks at ∼0.85 and 1.0 are present in the distribution, signaling the inhomogeneous distribution of $V^{3+}$ and $V^{4+}$. The horizontal dashed line in gray indicates the statistically averaged ratio, which is close to 0.9 for both films.

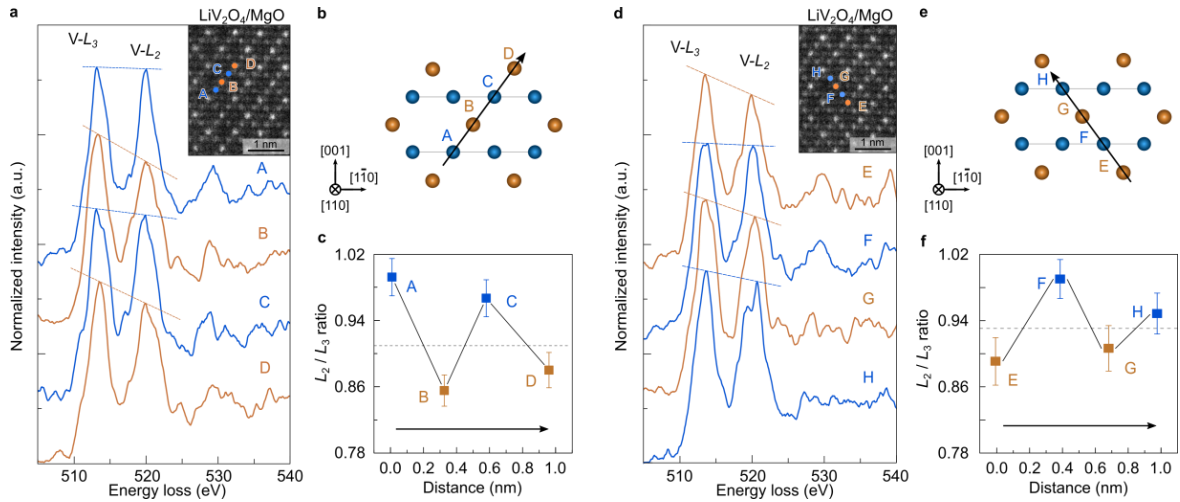

**Fig. 4 | Charge ordering of V ions in the strained film of LiV$_2$O$_4$ on MgO. a,d,** V $L_{2,3}$ edge spectra of the strained LiV$_2$O$_4$ film, taken from the V sites from A to D along the [1$\bar{1}$2] direction and from the V sites from E to H along the [$\bar{1}$12] direction, respectively, in the inset where the STEM-HAADF images are shown. The spectra were processed through a power-law background subtraction followed by a normalization to the $L_3$ intensity maximum. The spectra are vertically offset for clarity. **b,e,** The V pyrochlore sublattice viewed along the [110] direction, showing yellow V$^{3+}$ chains along the [110] direction and blue V$^{4+}$ chains along the [1$\bar{1}$0] direction (See also Figs. 1a and 1b). The black arrows indicate the line-scan direction. **c,f,** The $L_2/L_3$ ratio profiles in space obtained from the V $L_{2,3}$ spectra in **a** and **d**, showing the alternating V$^{3+}$ and V$^{4+}$ arrangement. The error bar originates from the standard error in the estimate of V $L_{2,3}$ intensity (maximum value).

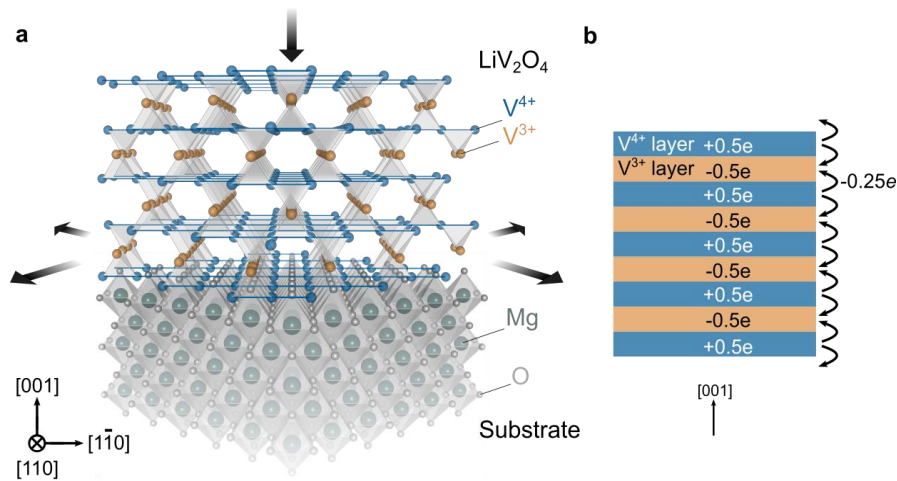

**Fig. 5 | Verwey-type charge ordering (CO) in the strained LiV$_2$O$_4$ film on MgO. a,** The charge-ordered structure of the epitaxially strained LiV$_2$O$_4$ film on MgO. The black arrows show the out-of-plane compressive and in-plane tensile strains produced by the interface with the MgO substrate. The V$^{3+}$ (yellow) and V$^{4+}$ (blue) chains run along the [110] and [1$\bar{1}$0] directions, respectively, giving rise to an alternating stack of V$^{3+}$ and V$^{4+}$ layers along [001]. **b,** As the CO takes place, electric charge is redistributed to the V$^{3+}$ and V$^{4+}$ layer with an excess charge −0.5$e$ amd +0.5$e$, respectively, from the spatially homogeneous V$^{+3.5}$ state in the unstrained LiV$_2$O$_4$. This leads to inter- and intra-layer Coulomb interactions becoming attractive and repulsive, which energetically favors the contraction and the expansion of the lattice along the out-of-plane and the in-plane directions, respectively. The preferred deformation is consistent with the strain-induced lattice deformation depicted in **a**.

Table I. Substrate parameters and mismatch to LiV$_2$O$_4$.

| Substrate | Orientation | $a$ (Å) | Mismatch to $a_{LVO}/2$ (%) |
|---|---|---|---|
| STO | (001) | 3.905 | −5.2 |
| MgO | (001) | 4.211 | +2.2 |
| LiV$_2$O$_4$ | - | 8.240 | - |

**References**

1. Kondo, S. *et al.* LiV$_2$O$_4$: a heavy fermion transition metal oxide. *Phys. Rev. Lett.* **78**, 3729-3732, doi:10.1103/PhysRevLett.78.3729 (1997).
2. Urano, C. *et al.* LiV$_2$O$_4$ spinel as a heavy-mass fermi liquid: anomalous transport and role of geometrical frustration. *Phys. Rev. Lett.* **85**, 1052-1055, doi:10.1103/PhysRevLett.85.1052 (2000).
3. Anisimov, V. I. *et al.* Electronic structure of the heavy fermion metal LiV$_2$O$_4$. *Phys. Rev. Lett.* **83**, 364-367, doi:10.1103/PhysRevLett.83.364 (1999).
4. Hopkinson, J. & Coleman, P. LiV$_2$O$_4$: frustration induced heavy fermion metal. *Phys. Rev. Lett.* **89**, 267201, doi:10.1103/PhysRevLett.89.267201 (2002).
5. Arita, R., Held, K., Lukoyanov, A. & Anisimov, V. Doped Mott insulator as the origin of heavy-fermion behavior in LiV$_2$O$_4$. *Phys. Rev. Lett.* **98**, 166402, doi:10.1103/PhysRevLett.98.166402 (2007).
6. Jönsson, P. E. *et al.* Correlation-driven heavy-fermion formation in LiV$_2$O$_4$. *Phys. Rev. Lett.* **99**, 167402, doi:10.1103/PhysRevLett.99.167402 (2007).
7. Fulde, P., Yaresko, A. N., Zvyagin, A. A. & Grin, Y. On the origin of heavy quasiparticles in LiV$_2$O$_4$. *EPL* **54**, 779-785, doi:10.1209/epl/i2001-00322-3 (2001).
8. Reuter, B. & Jaskowsky, J. LiV$_2$O$_4$–MgV$_2$O$_4$, neues Spinellsystem mit Valenzhalbleitereigenschaften. *Angew. Chem.* **72**, 209, doi:10.1002/ange.19600720610 (1960).
9. Shimizu, Y. *et al.* An orbital-selective spin liquid in a frustrated heavy fermion spinel LiV$_2$O$_4$. *Nat. Commun.* **3**, 1-5, doi:10.1038/ncomms1979 (2012).
10. Pauling, L. The structure and entropy of ice and of other crystals with some randomness of atomic arrangement. *J. Am. Chem. Soc.* **57**, 2680-2684, doi:10.1021/ja01315a102 (1935).
11. Anderson, P. W. Ordering and antiferromagnetism in ferrites. *Phys. Rev.* **102**, 1008-1013, doi:10.1103/PhysRev.102.1008 (1956).
12. Pinsard-Gaudart, L. *et al.* Pressure-induced structural phase transition in LiV$_2$O$_4$. *Phys. Rev. B* **76**, 045119, doi:10.1103/PhysRevB.76.045119 (2007).
13. Fujiwara, K., Miyoshi, K., Takeuchi, J., Shimaoka, Y. & Kobayashi, T. Li NMR in LiV$_2$O$_4$ under high pressure. *J. Phys. Condens. Matter* **16**, S615, doi:10.1088/0953-8984/16/11/007 (2004).
14. Browne, A. J., Pace, E. J., Garbarino, G. & Attfield, J. P. Structural study of the pressure-induced metal-insulator transition in LiV$_2$O$_4$. *Phys. Rev. Mater.* **4**, 015002, doi:10.1103/PhysRevMaterials.4.015002 (2020).
15. Takeda, K. *et al.* Pressure-induced charge ordering of LiV$_2$O$_4$. *Physica B Condens. Matter* **359**, 1312-1314, doi:10.1016/j.physb.2005.01.390 (2005).
16. Irizawa, A. *et al.* Direct observation of a pressure-induced metal-insulator transition in LiV$_2$O$_4$ by optical studies. *Phys. Rev. B* **84**, 235116, doi:10.1103/PhysRevB.84.235116 (2011).
17. Hwang, H. Y. *et al.* Emergent phenomena at oxide interfaces. *Nat. Mater.* **11**, 103-113, doi:10.1038/nmat3223 (2012).
18. Ramesh, R. & Schlom, D. G. Creating emergent phenomena in oxide superlattices. *Nat. Rev. Mater.* **4**, 257-268, doi:10.1038/s41578-019-0095-2 (2019).
19. Zubko, P., Gariglio, S., Gabay, M., Ghosez, P. & Triscone, J.-M. Interface physics in complex oxide heterostructures. *Annu. Rev. Condens. Matter Phys.* **2**, 141-165, doi:10.1146/annurev-conmatphys-062910-140445 (2011).
20. Okada, Y. *et al.* Scanning tunnelling spectroscopy of superconductivity on surfaces of LiTi$_2$O$_4$ (111) thin films. *Nat. Commun.* **8**, 1-7, doi:10.1038/ncomms15975 (2017).
21. Jin, K. *et al.* Anomalous magnetoresistance in the spinel superconductor LiTi$_2$O$_4$. *Nat. Commun.* **6**, 1-8, doi:10.1038/ncomms8183 (2015).
22. Hu, W. *et al.* Emergent superconductivity in single-crystalline MgTi$_2$O$_4$ films via structural engineering. *Phys. Rev. B* **101**, 220510, doi:10.1103/PhysRevB.101.220510 (2020).

23   Chopdekar, R. V., Wong, F. J., Takamura, Y., Arenholz, E. & Suzuki, Y. Growth and characterization of superconducting spinel oxide LiTi$_2$O$_4$ thin films. *Physica C Supercond.* **469**, 1885-1891, doi:10.1016/j.physc.2009.05.009 (2009).
24   Yajima, T. *et al.* Heavy-fermion metallic state and Mott transition induced by Li-ion intercalation in LiV$_2$O$_4$ epitaxial films. *Phys. Rev. B* **104**, 245104, doi:10.1103/PhysRevB.104.245104 (2021).
25   Chmaissem, O., Jorgensen, J. D., Kondo, S. & Johnston, D. C. Structure and thermal expansion of LiV$_2$O$_4$: correlation between structure and heavy fermion behavior. *Phys. Rev. Lett.* **79**, 4866-4869, doi:10.1103/PhysRevLett.79.4866 (1997).
26   Poisson, S. D. Note sur l'extension des fils et des plaques élastiques. *Ann. Chim. Phys* **36**, 384-387 (1827).
27   Kadowaki, K. & Woods, S. Universal relationship of the resistivity and specific heat in heavy-fermion compounds. *Solid State Commun* **58**, 507-509, doi:10.1016/0038-1098(86)90785-4 (1986).
28   Johnston, D., Swenson, C. & Kondo, S. Specific heat (1.2–108 K) and thermal expansion (4.4–297 K) measurements of the 3$d$ heavy-fermion compound LiV$_2$O$_4$. *Phys. Rev. B* **59**, 2627, doi:10.1103/PhysRevB.59.2627 (1999).
29   Vecchio, I. L. *et al.* Optical conductivity of V$_4$O$_7$ across its metal-insulator transition. *Phys. Rev. B* **90**, 115149, doi:10.1103/PhysRevB.90.115149 (2014).
30   Sparrow, T., Williams, B., Rao, C. & Thomas, J. $L_3$/$L_2$ white-line intensity ratios in the electron energy-loss spectra of 3$d$ transition-metal oxides. *Chem. Phys. Lett.* **108**, 547-550, doi:10.1016/0009-2614(84)85051-4 (1984).
31   Kalavathi, S., Amirthapandian, S., Chandra, S., Sahu, P. C. & Sahu, H. Valence state, hybridization and electronic band structure in the charge ordered AlV$_2$O$_4$. *J. Phys. Condens. Matter* **26**, 015601, doi:10.1088/0953-8984/26/1/015601 (2013).
32   Lin, X. W., Wang, Y. Y., Dravid, V. P., Michalakos, P. M. & Kung, M. C. Valence states and hybridization in vanadium oxide systems investigated by transmission electron-energy-loss spectroscopy. *Phys. Rev. B* **47**, 3477-3481, doi:10.1103/PhysRevB.47.3477 (1993).
33   Verwey, E. J. W. Electronic Conduction of magnetite (Fe$_3$O$_4$) and its transition point at low temperatures. *Nature* **144**, 327-328, doi:10.1038/144327b0 (1939).
34   Browne, A. J., Kimber, S. A. & Attfield, J. P. Persistent three-and four-atom orbital molecules in the spinel AlV$_2$O$_4$. *Phys. Rev. Mater.* **1**, 052003, doi:10.1103/PhysRevMaterials.4.015002 (2017).
35   Senn, M. S., Wright, J. P. & Attfield, J. P. Charge order and three-site distortions in the Verwey structure of magnetite. *Nature* **481**, 173-176, doi:10.1038/nature10704 (2012).

# Supplementary information for "Crystallization of heavy fermions via epitaxial strain in spinel $LiV_2O_4$ thin film"

U. Niemann[1,2*], Y.-M. Wu[1*], R. Oka[1,2], D. Hirai[2], Y. Wang[1], Y. E. Suyolcu[1], M. Kim[1], P. A. van Aken[1], and H. Takagi[1,2,3]

[1]Max Plank Institute for Solid State Research, Heisenbergstraße 1, 70569 Stuttgart, Germany

[2]Department of Physics, University of Tokyo, Hongo 7-3-1, Bunkyo-ku, Tokyo 113-0033, Japan

[3]Insitute for Functional Matter and Quantum Technologies, University of Stuttgart, Pfaffenwaldring 57, 70550 Stuttgart, Germany.

*These authors contributed equally to this work.

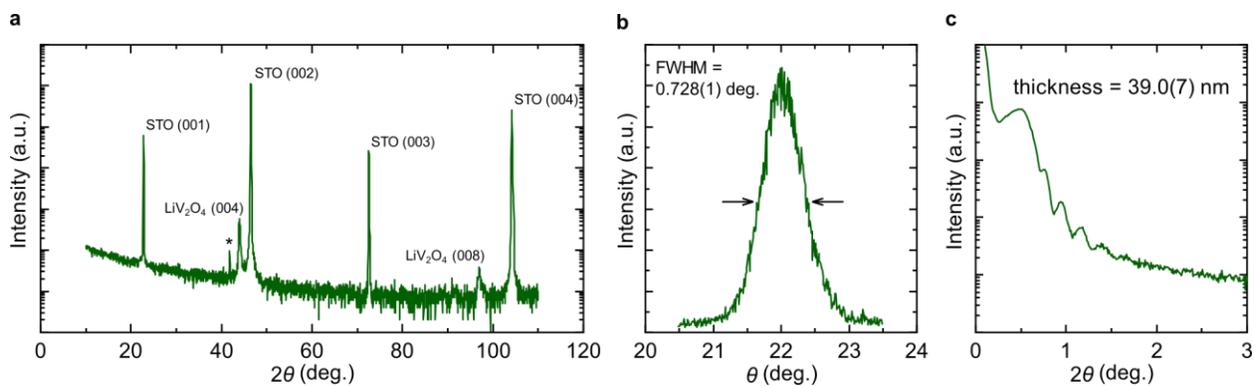

**Fig. S1 XRD characterization of a LiV$_2$O$_4$ thin film on STO using a Cu $K_α$ source. a,** X-ray $θ$-$2θ$ scan of a LiV$_2$O$_4$ thin film on STO. The peak with the asterisk originates from the STO (002) reflection due to a minor Cu $K_β$ component. deg. and a.u. denote degree and arbitrary unit, respectively. **b,** Rocking curve of the LiV$_2$O$_4$ (004) peak. The estimated full width at half maximum (FWHM) using a Pseudo-Voigt function is 0.728(1) deg. **c,** X-ray reflectivity profile. From Kiessig oscillations, the film thickness is estimated to be 39.0(7) nm.

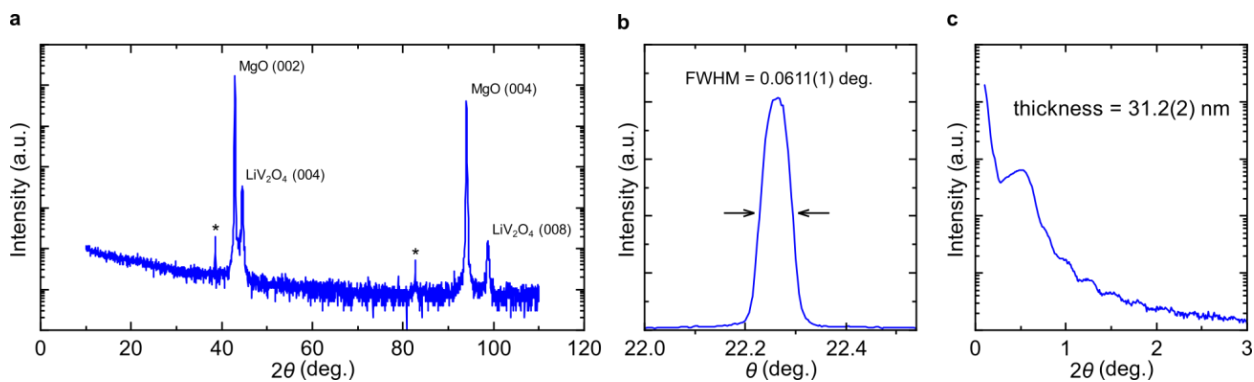

**Fig. S2 XRD characterization of a LiV$_2$O$_4$ thin film on MgO using a Cu $K_α$ source. a,** X-ray $θ$-$2θ$ scan of a LiV$_2$O$_4$ thin film on MgO. The peaks with the asterisk originate from the MgO (002) and (004) reflections due to a minor Cu $K_β$ component. **b,** Rocking curve of the LiV$_2$O$_4$ (004) peak. The estimated FWHM using a Pseudo-Voigt function is 0.0611(1) deg. **c,** X-ray reflectivity profile. The film thickness is estimated to be 31.2(2) nm.

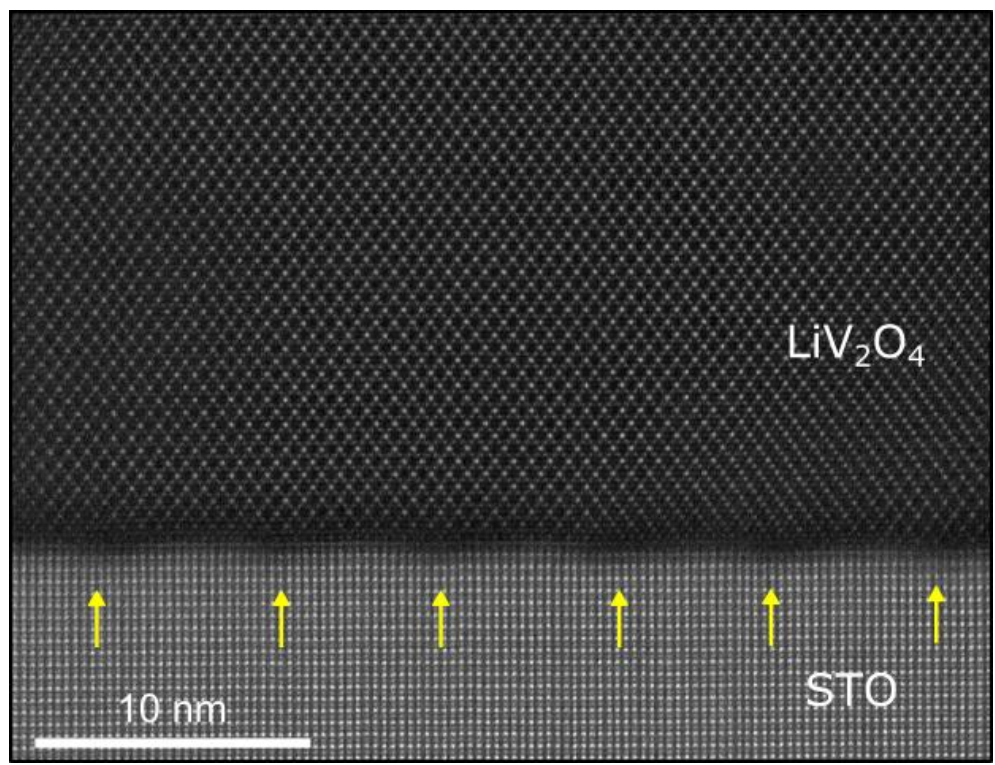

**Fig. S3** Representative STEM-HAADF image of a LiV$_2$O$_4$ film on STO showing periodic misfit dislocations at the interface, which are indicated by the yellow arrows.

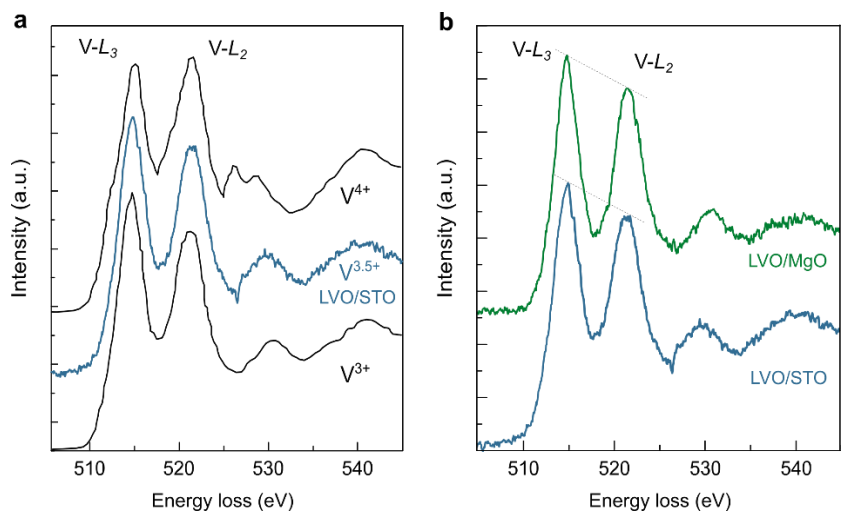

**Fig. S4 a,** Vanadium $L_{2,3}$ spectrum for the $LiV_2O_4$ film on STO averaged over all the spatially resolved data in Fig. 3b (blue). Two reference $L_{2,3}$ spectra for $V^{3+}$ and $V^{4+}$ obtained from the literature on $V_2O_3$ and $VO_2$ (ref. [1]) are shown for comparison (black). The averaged LVO/STO spectrum appears as a 1:1 superposition of the $V^{3+}$ and $V^{4+}$ spectra: the $L_2/L_3$ ratio is ~0.9, close to an average of 0.83 for $V^{3+}$ and 1.02 for $V^{4+}$. LVO denotes $LiV_2O_4$. **b,** Comparison of the averaged spectra of the LVO on STO (blue) and MgO (green) in Figs. 3b and 3d, showing an almost identical V $L_{2,3}$ edge fine structure. The spectra are normalized to the $L_3$ edge intensity maximum, and vertically offset for clarity.

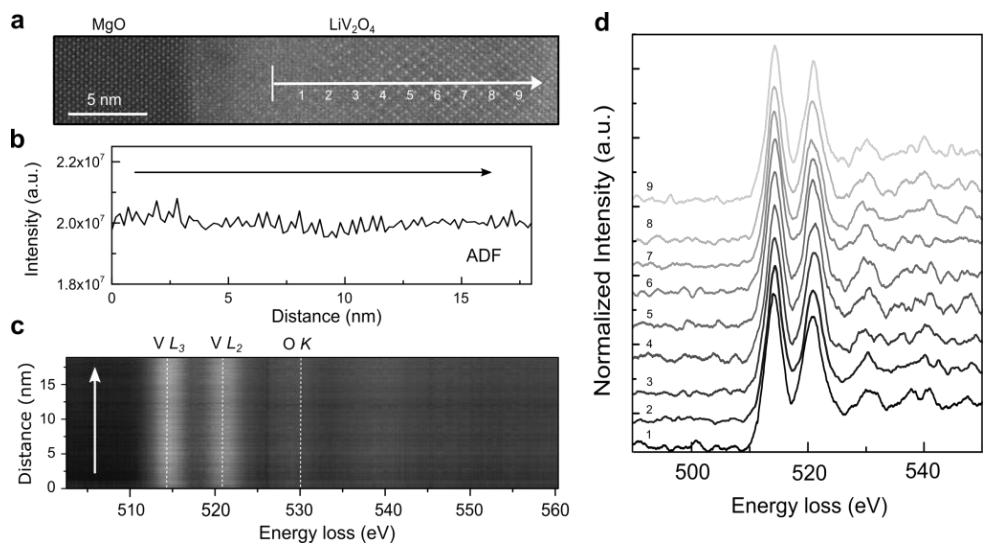

**Fig. S5 a,** STEM-HAADF image of the LiV$_2$O$_4$ film on MgO. The white arrow indicates the line of scanning on which the EELS spectra in **d** were acquired. The vertical line indicates the averaging width while scanning. **b,** Simultaneously recorded annular dark-field (ADF) intensity profile does not show any appreciable changes in the intensity, ensuring no electronic modification under the beam. **c,** Evolution of background-subtracted EELS line-scan spectra along the direction of the white arrow are shown in a gray scale, which indicate the two peaks at 513 and 521 eV corresponding to the V $L_3$ and $L_2$ white-lines, respectively. **d,** V $L_{2,3}$ spectra collected from positions 1-9 in **a**. Each spectrum is normalized to the $L_3$ edge intensity maximum, and vertically offset for clarity.

**Supplementary Note 1: EELS data processing**

The 3$d$ states of V are probed via the V $L_{2,3}$ edges at 513 and 521 eV. A simple power law background subtraction was used and the multiple scattering was alleviated to minimize the effect of thickness for the $L_2/L_3$ ratio calculations as available in Digital Micrograph. No principal components analysis (PCA) or other noise-reducing data processing were employed. Due to the limited energy resolution (~0.7 eV) governed by the energy dispersion that we chose for a sufficiently high spectral signal-to-noise ratio, the delocalization of the inelastic electron scattering, as well as the experimental difficulty of calibrating the energy scales, we do not leverage the chemical shift but rather focus on the changes in the $L_2/L_3$ intensity ratio to determine the V valence. The V valence was determined by using the maximum intensities of the two peaks $I(L_2)/I(L_3)$ using the method of Ref. 2, since the continuum contribution cannot be removed due to the overlap with the O $K$ edge at ~532 eV.

For quantitative analysis, reference spectra for $V^{3+}$ and $V^{4+}$ from the literature[1] were used to confirm the expected V valence state of the LiV$_2$O$_4$ films, as shown in Fig. S4a. The $L_2/L_3$ ratios are 0.83 and 1.02 for $V^{3+}$ and $V^{4+}$ spectra, respectively. As shown in Fig. S4b, both LiV$_2$O$_4$ films on STO and MgO show the same averaged $L_2/L_3$ ratio of 0.9 and almost identical V $L_{2,3}$ and O $K$ edge fine structures, consistent with the expected the $V^{3.5+}$ formal valence in LiV$_2$O$_4$. This indicates that the films are free from radiation damage and any resultant O non-stoichiometry. We note that the potential intermixing of an inelastic EELS signal due to a delocalization effect could underestimate the valence states[3,4].

**References**

[1] Lin, X. W. *et al.* Valence states and hybridization in vanadium oxide systems investigated by transmission electron-energy-loss spectroscopy. *Phys. Rev. B* **47**, 3477-3481 (1993).

[2] Riedl, T. *et al.* Extraction of EELS white-line intensities of manganese compounds: Methods, accuracy, and valence sensitivity. *Ultramicroscopy* **106**, 284-291 (2006).

[3] Botton, G. *et al.* Quantification of the EELS near-edge structures to study Mn doping in oxides. *J. Microsc.* **180**, 211–216 (1995).

[4] Egerton, R. F. Electron energy-loss spectroscopy in the electron microscope (Springer Science & Business Media, 2011).